\documentclass[review]{elsarticle}

\usepackage[margin=1in]{geometry}
\usepackage{float}
\usepackage{lineno,hyperref}
\usepackage{color}
\usepackage{amssymb}
\usepackage{graphics}
\usepackage{graphicx}
\usepackage{pbox}
\usepackage{mhchem}
\usepackage{xcolor} 
\usepackage[ampersand]{easylist}
\usepackage{lineno}
\usepackage{booktabs}
\usepackage{multirow}
\usepackage{subcaption}
\usepackage{amsmath}
\usepackage{paralist}
\usepackage{soul}
\usepackage{siunitx}
\usepackage{amsmath}
\usepackage{url}
\usepackage{makecell}
\usepackage{hyperref}

\newcommand{\argmin}{\mathop{\mathrm{argmin}}\limits}
\newcommand{\solve}{\mathop{\text{Solve}}\limits}
\newcommand{\cm}{\text{cm}} 
\newcommand{\g}{\text{g}}
 
\newcommand{\eff}{\text{eff}}

\newcommand{\PE}{\text{PE}}
\newcommand{\CS}{\text{CS}}
\newcommand{\PP}{\text{PP}}
\newcommand{\appropto}{\mathrel{\vcenter{
  \offinterlineskip\halign{\hfil$##$\cr
    \propto\cr\noalign{\kern2pt}\sim\cr\noalign{\kern-2pt}}}}}

\hypersetup{  
    colorlinks=true,
    linkcolor=blue,
    filecolor=magenta,      
    urlcolor=cyan,
    pdftitle={Sharelatex Example},
    bookmarks=true,
    pdfpagemode=FullScreen,
    citecolor=blue
    }

\modulolinenumbers[1]

\makeatletter
\def\ps@pprintTitle{%
 \let\@oddhead\@empty
 \let\@evenhead\@empty
 \def\@oddfoot{\centerline{\thepage}}%
 \let\@evenfoot\@oddfoot}
\makeatother

\begin{document}

\begin{frontmatter}

\title{Fundamental Limitations of Dual Energy X-ray Scanners for Cargo Content Atomic Number Discrimination}

\author[MITaddress]{Peter Lalor\corref{corauthor}}
\cortext[corauthor]{Corresponding author 
\\\hspace*{13pt} Email address: plalor@mit.edu
\\\hspace*{13pt} Telephone: (925) 453-1876 
\\\hspace*{13pt} 138 Cherry St, Cambridge, MA 02139}

\author[MITaddress]{Areg Danagoulian}

\address[MITaddress]{Department of Nuclear Science and Engineering, Massachusetts Institute of Technology, Cambridge, MA 02139, USA}

\begin{abstract}
To combat the risk of nuclear smuggling, radiography systems are deployed at ports to scan cargo containers for concealed illicit materials. Dual energy radiography systems enable a rough elemental analysis of cargo containers due to the $Z$-dependence of photon attenuation, allowing for improved material detection. This work studies the capabilities for atomic number discrimination using dual energy MeV systems by considering dual energy $\{6, 4\}$ MeV, $\{10, 6\}$ MeV, and $\{10, 4\}$ MeV bremsstrahlung beams. Results of this analysis show that two different pure materials can sometimes produce identical transparency measurements, leading to a fundamental ambiguity when differentiating between materials of different atomic numbers. Previous literature has observed this property, but the extent of the limitation is poorly understood and the cause of the degeneracy is generally inadequately explained. This non-uniqueness property stems from competition between photoelectric absorption and pair production and is present even in systems with perfect resolution and zero statistical noise. These findings are validated through Monte Carlo transparency simulations. Results of this study show that currently deployed commercial radiographic systems are fundamentally incapable of distinguishing between high-$Z$ nuclear materials and miscellaneous mid-$Z$ cargo contents. 
\end{abstract}
\end{frontmatter}
\begin{sloppypar}
\twocolumn


\section{Introduction}
\label{Introduction}

Since 1993, the International Atomic Energy Agency (IAEA) database shows 320 incidents of theft or illicit tracking involving nuclear material, 20 of which involve highly enriched uranium (HEU) or plutonium~\cite{IAEA2022}. Economic costs of a nuclear detonation at a U.S. port could exceed \$1 trillion~\cite{Meade2006}, while a smaller scale radiological dispersal device (``dirty bomb") could results in losses of tens of billions due to trade disruption and port shutdown costs~\cite{Rosoff2007}. In order to combat this threat, the U.S. Congress passed the SAFE Port Act in 2006, which mandated 100 percent screening of U.S. bound cargo and 100 percent scanning of high-risk containers~\cite{PLAW109-347}. This act also called for a full-scale implementation to scan all inbound containers at foreign ports prior to U.S. entry. These measures would inspect cargo for special nuclear materials (SNM), which a terrorist might attempt to smuggle through U.S. ports in order to build and subsequently detonate a nuclear bomb~\cite{Cochran2008}.

During the screening of cargo at U.S. ports, cargo passes by a radiation portal monitor (RPM), which detects neutron and/or gamma radiation that is passively being emitted by nuclear and radiological materials that may be hidden in a cargo container~\cite{Kouzes2005, Kouzes2008}. If the measurement is significantly above background measurements, an alarm would trigger, detecting the smuggling attempt. However, a smuggler could defeat this mode of detection through sufficient shielding of the smuggled material~\cite{Gaukler2010, Miller2015}.

Non-intrusive inspection (NII) technologies further enable the detection of illicit materials~\cite{NII}. These radiography scanners use X-rays or gamma rays, which are directed through the cargo and measured by a detector downstream from the container~\cite{Rapiscan_X-Ray}. The attenuation of these photons gives a sensitive density image of the scanned cargo. Dual energy X-ray systems could provide a rough elemental analysis of the cargo contents, since the attenuation of X-rays is sensitive to the atomic number of the material. High-$Z$ classification, which the domestic nuclear detection office (DNDO) defines as $Z \geq 72$~\cite{Perticone2010}, would be particularly useful for detecting smuggled nuclear materials and high-$Z$ shielding such as lead or tungsten. 

\section{Background}
\label{Background}

When a radiography system scans a material of area density $\lambda$ and atomic number $Z$, it measures the transparency of the photon beam, defined as the detected charge in the presence of the material normalized by the open beam measurement. Using the Beer-Lambert law, we can express the photon transparency $T(\lambda, Z)$ as follows:

\begin{equation}
T(\lambda, Z) = \frac{\int_0^{\infty} D(E) \phi(E) e^{-\mu (E, Z) \lambda} dE}{\int_0^{\infty}D(E) \phi(E) dE}
\label{transparency}
\end{equation}

where $\mu(E, Z)$ is the mass attenuation coefficient, $\phi(E)$ is the differential photon beam spectrum, and $D(E)$ is the detector response function. Equation \ref{transparency} assumes only noninteracting photons are detected, and thus ignores the effects of scattered radiation. Our past work found that these secondary effects can be accurately captured by replacing the mass attenuation coefficient in Eq. \ref{transparency} with a semiempirical mass attenuation coefficient $\tilde \mu(E, Z)$~\cite{Lalor2023}:

\begin{equation}
\tilde \mu(E, Z) = a\mu_\PE(E, Z) + b\mu_\CS(E, Z) + c\mu_\PP(E, Z)
\label{semiempirical_mass_atten}
\end{equation}

where $a$, $b$, and $c$ are determined through a calibration step, as described in Section \ref{calculating_abc}. In Eq. \ref{semiempirical_mass_atten}, $\mu_\PE(E, Z)$, $\mu_\CS(E, Z)$, and $\mu_\PP(E, Z)$ are the mass attenuation coefficients from the photoelectric effect (PE), Compton scattering (CS), and pair production (PP), calculated from NIST cross section tables~\cite{NIST}. Each of these mass attenuation coefficients depend on $Z$ in different ways~\cite{Leo1994}:

\begin{align}
\mu_\PE(E, Z) &\appropto \frac{Z^{4-5}}{A} f_\PE(E) \label{muPE} \\
\mu_\CS(E, Z) &\appropto \frac{Z}{A} f_\CS(E) \label{muCS} \\
\mu_\PP(E, Z) &\appropto \frac{Z^2}{A} f_\PP(E) \label{muPP}
\end{align}

where $A$ is the atomic mass of the material and $f(E)$ is the energy dependence of the corresponding interaction. The photoelectric effect is the dominant interaction mechanism at lower energies, Compton scattering at intermediate energies, and pair production at higher energies.

Using the subscripts $\{H, L\}$ to refer to the \{high, low\} energy beam, the dual energy principle can be represented as a system of two equations and two unknowns through Eq. \ref{2x2eq}:

\begin{equation}
\hat \lambda, \hat Z = \solve\limits_{\lambda, Z}
\begin{cases} 
      T_H(\lambda, Z) = T_H \\
      T_L(\lambda, Z) = T_L
\end{cases}
\label{2x2eq}
\end{equation}

where $\{T_H, ~T_L\}$ are the measured transparencies and $\{T_H(\lambda, Z), ~T_L(\lambda, Z)\}$ are calculated by Eq. \ref{transparency} using the semiempirical mass attenuation coefficient (Eq. \ref{semiempirical_mass_atten}). In practice, Eq. \ref{2x2eq} is typically solved using a precomputed reverse 2D lookup table~\cite{Zhang2005}.

\section{Analysis}
\label{Analysis}

\subsection{Methodology}
\label{Methodology}

The ability for a dual energy system to reconstruct the atomic number of an imaged object is determined by the properties of Eq. \ref{2x2eq}. To study the capabilities and limitations of different possible dual energy systems, this work will analyze dual energy $\{6, 4\}$ MeV bremsstrahlung beams, dual energy $\{10, 6\}$ MeV bremsstrahlung beams, and dual energy $\{10, 4\}$ MeV bremsstrahlung beams.

Bremsstrahlung beams are frequently used for cargo radiography applications due to their availability, flexibility, and high photon output~\cite{Martz2017}. The $\{6, 4\}$ MeV analysis case is motivated by its prevalence in existing commercial systems, such as the Rapiscan Eagle R60${}^\text{\textregistered}$ rail cargo inspection system~\cite{Rapiscan_overview}. The $\{10, 6\}$ MeV analysis case explores the benefits of using higher energy beams, with the upper $10$ MeV cutoff chosen to limit safety concerns associated with neutron production and residual activation~\cite{Chen2007}. The $\{10, 4\}$ MeV analysis case examines the benefits of improving beam contrast by simultaneously maximizing pair production in the high energy beam and minimizing pair production in the low energy beam.

For this analysis, characteristic $4$ MeV, $6$ MeV, and $10$ MeV bremsstrahlung beam spectra were simulated using a representative geometry based on the work of Henderson~\cite{Henderson2019}. The simulation geometry is detailed in Section \ref{Simulation}, and the resulting photon spectra are shown in Fig. \ref{spectra}. While the beam spectra used in this study are far more simplified than those in real cargo security applications, the analysis to be performed in Section \ref{Analysis} applies generally.

\subsection{Regions with Non-unique Solutions}
\label{non-unique solutions}

Novikov introduces the $\alpha$-curve as a method for visualizing the solution space to Eq. \ref{2x2eq}~\cite{Novikov1999}. We first make a log transform of Eq. \ref{transparency} to define a new variable:

\begin{equation}
\alpha(\lambda, Z) = -\log T(\lambda, Z) \\
\label{alpha}
\end{equation}

\begin{figure}
\begin{centering}
\includegraphics[width=0.49\textwidth]{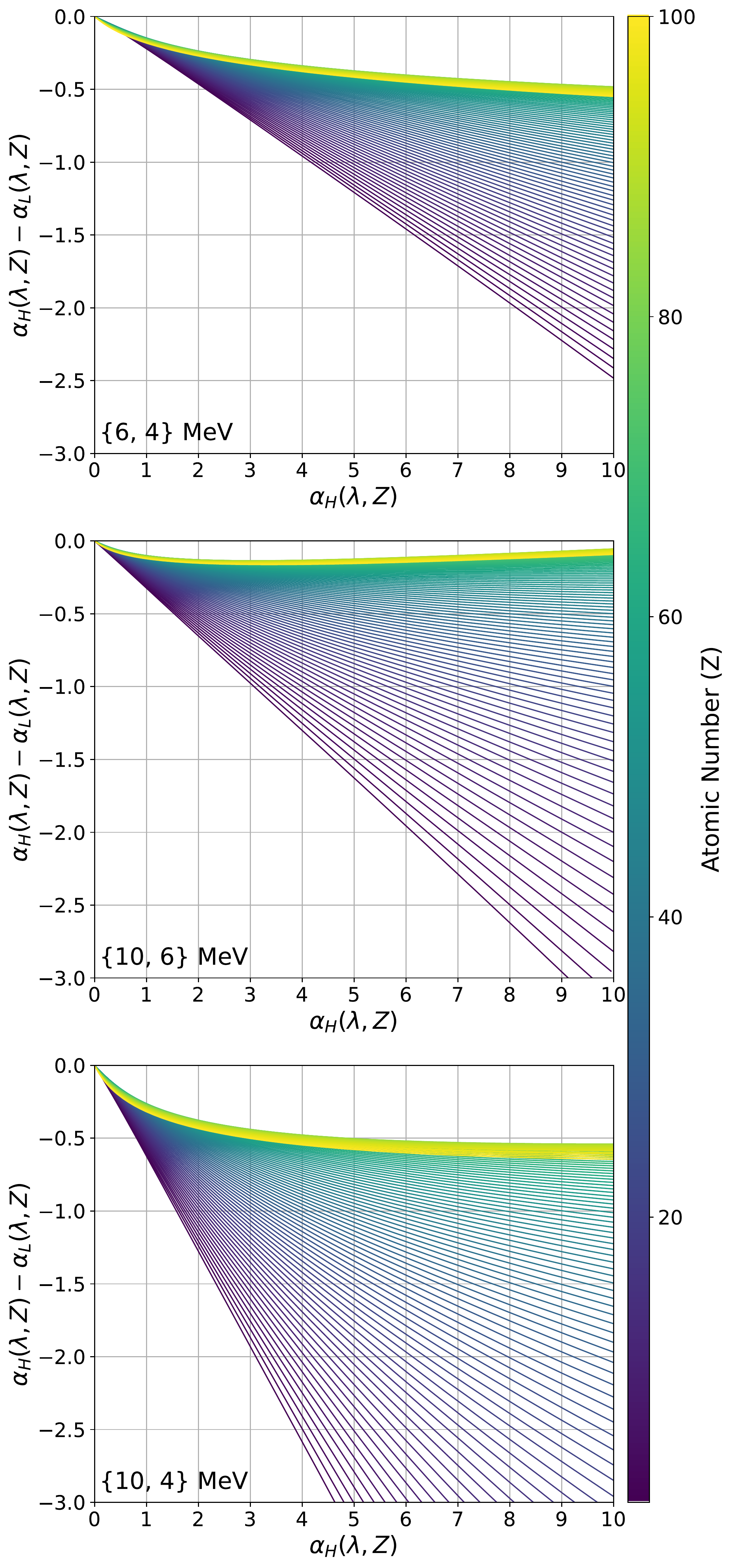}
\caption{$\alpha$-curves for the three analysis cases described in Section \ref{Methodology}, generated using dual energy $\{6, 4\}$ MeV (top), $\{10, 6\}$ MeV (center), and $\{10, 4\}$ MeV (bottom). The transparencies are calculated using Eq. \ref{transparency}, and $\alpha_H(\lambda, Z)$ and $\alpha_L(\lambda, Z)$ are calculated from the transparencies using Eq. \ref{alpha}. Materials of different atomic numbers create different $\alpha$-lines, and the position on each $\alpha$-line is determined by the area density $\lambda$ of the material (with the top left corner corresponding to $\lambda=0$). The $\alpha$-curve shows the relative behavior of $\alpha_H$ and $\alpha_L$ across different elements as material thickness is varied.}
\label{alpha_curve}
\end{centering}
\end{figure}

Figure \ref{alpha_curve} shows an $\alpha$-curve for each of the three analysis cases described in Section \ref{Methodology}. An $\alpha$-curve is a plot of $\alpha_H(\lambda, Z) - \alpha_L(\lambda, Z)$ versus $\alpha_H(\lambda, Z)$ for different elements and area densities. In this way, every element corresponds to a single line on the $\alpha$-curve (henceforth referred to as an $\alpha$-line), and the position on the $\alpha$-line is determined by the area density of the material. The shape and structure of the $\alpha$-curve depends on the implicit parameters of Eq. \ref{transparency}, i.e. the high- and low-energy incident beam spectra $\phi_H (E)$ and $\phi_L (E)$ and the detector response function $D(E)$.

Overlapping $\alpha$-lines on the $\alpha$-curve correspond to regions with non-unique solutions to Eq. \ref{2x2eq}. In other words, there may exist a pair of solutions $(\lambda_1, Z_1)$ and $(\lambda_2, Z_2)$ which both solve Eq. \ref{2x2eq} for transparency measurements $\{T_H, T_L\}$, creating a fundamental ambiguity as to which atomic number solution to choose. It is worth emphasizing that even a system with perfect resolution and zero statistical noise would still be unable to differentiate between different elements in regions with non-unique solutions.

The presence of non-unique solutions was first observed by Novikov for heavy metals and thin materials~\cite{Novikov1999}. However, the fundamental physics describing this phenomenon is generally poorly explained in existing literature, leading to misconceptions as to the source of the ambiguity. The presence of non-unique solutions stems from competition between the photoelectric effect and pair production. Pair production interactions are more common in the high energy beam, and the $\frac{Z^2}{A}$ dependence of $\mu_\PP(E, Z)$ (Eq. \ref{muPP}) provides the mathematical basis for dual-energy radiography as a means of atomic number discrimination. However, photoelectric interactions are more common in the low energy beam, which become more significant at high-$Z$ due to the strong $\frac{Z^{4-5}}{A}$ dependence of $\mu_\PE(E, Z)$ (Eq. \ref{muPE}). This results in a degeneracy in which a thicker low-$Z$ material and a thinner high-$Z$ material can produce the same high and low energy transparency measurements. This ambiguity is not present in low energy applications (such as medical imaging) due to the absence of pair production.

At low material area densities, this effect is most pronounced due to the significant low energy component of the bremsstrahlung spectra. The high probability of photoelectric interactions causes $\alpha$-lines to frequently overlap in this regime, resulting in atomic number ambiguity. At larger area densities, the bremsstrahlung beam hardens and the low energy component is selectively attenuated, significantly reducing the contribution from photoelectric absorption. In this regime, only high-$Z$ $\alpha$-lines overlap, limiting the non-uniqueness to heavy metals. This effect can be seen in the high-$Z$ regions of the $\alpha$-curves shown in Fig. \ref{alpha_curve} where after $Z \approx 80$, the $\alpha$-lines fold over, causing $\alpha$-lines with $Z \gtrsim 80$ to overlap other mid- to high-$Z$ $\alpha$-lines. This result rejects the intuition that the higher the material atomic number, the easier it will be to differentiate from other low-$Z$ or mid-$Z$ cargo. In other words, discriminating between $Z = 26$ and $Z = 94$ is more challenging than discriminating between $Z = 26$ and $Z = 82$, since the $Z=94$ $\alpha$-line is below the $Z = 82$ $\alpha$-line.

\begin{figure}
    \centering
    \begin{subfigure}[t]{0.50\textwidth}
        \includegraphics[width=\textwidth]{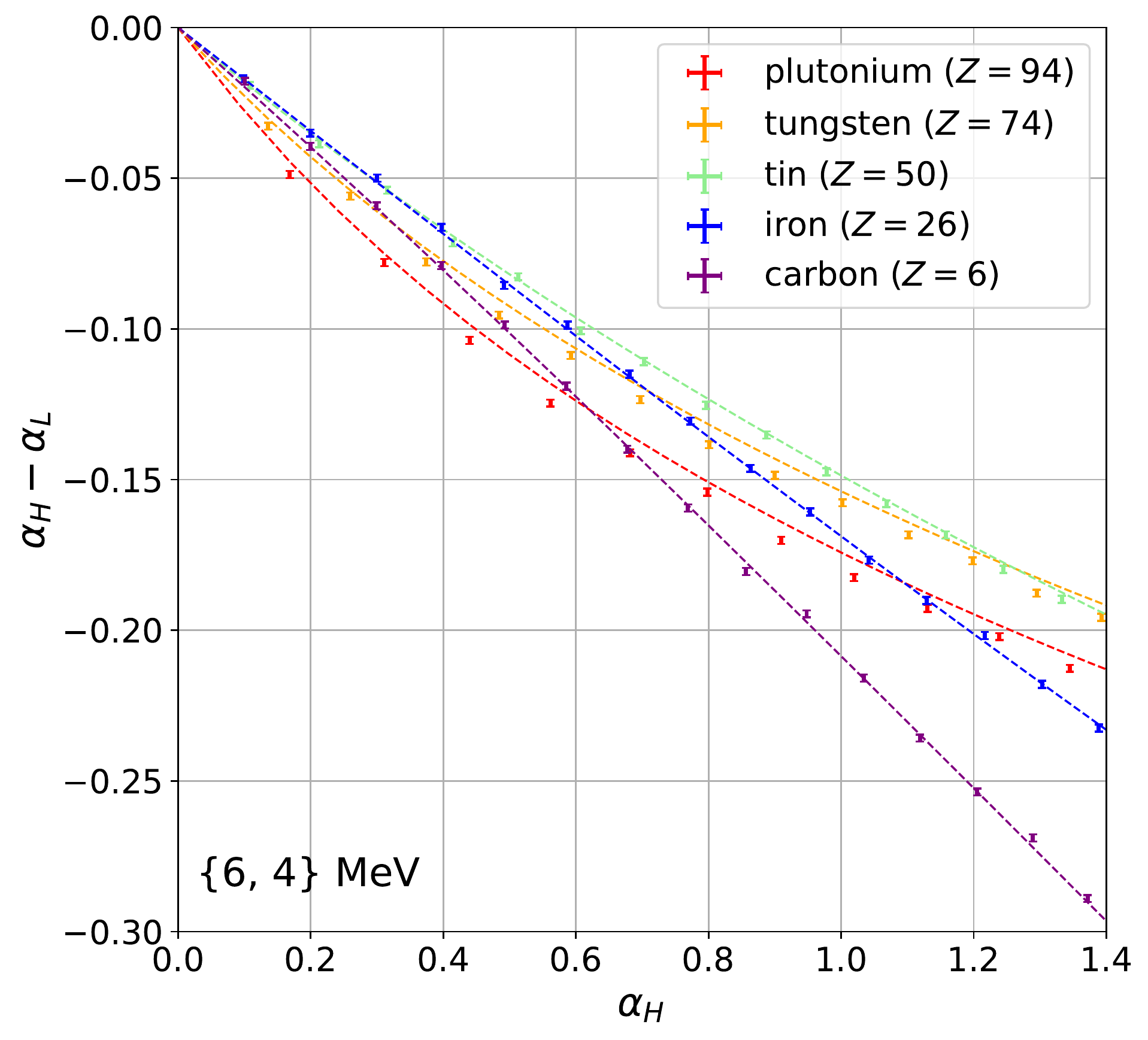}
        \caption{$\{6, 4\}$ MeV $\alpha$-curve, zoomed in at the low area density component. Different $\alpha$-lines frequently overlap, corresponding to non-unique atomic number solutions.}
        \label{validation1}
    \end{subfigure}
    \hfill
    \begin{subfigure}[t]{0.48\textwidth}
        \includegraphics[width=\textwidth]{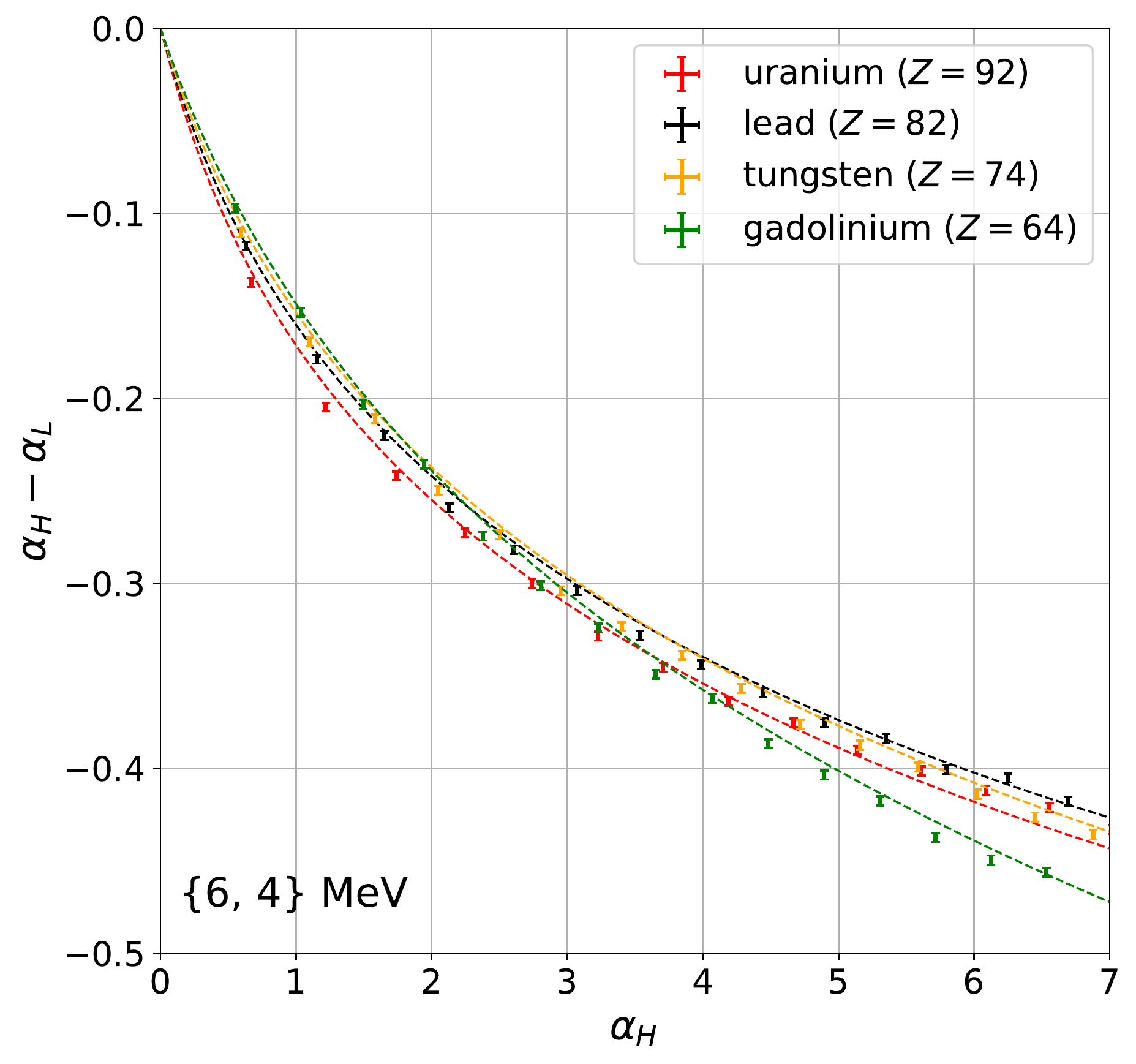}
        \caption{$\{6, 4\}$ MeV $\alpha$-curve only showing high-$Z$ elements. Material overlap occurs even for larger values of $\alpha$, demonstrating that the high-$Z$ degeneracy is still present for thick targets.}
    \label{validation2}
    \end{subfigure}
    \caption{Validating the non-uniqueness property described in Section \ref{non-unique solutions} using Monte Carlo simulations. Errorbars show simulation output, and dashed lines show theoretical output (Eq. \ref{transparency}).}
    \label{validation}
\end{figure}


To validate these results, transparency simulations were run in Geant4~\cite{Geant4, grasshopper}. $10$, $6$, and $4$ MeV bremsstrahlung beams were directed through different materials of known thicknesses, and the beam transparencies were recorded. The simulation geometry is described in more detail in Section \ref{Transparency Simulations}. Fig. \ref{validation} shows these simulated measurements on an $\alpha$-curve for the $\{6, 4\}$ MeV case, demonstrating that the theoretically predicted $\alpha$-line overlaps are present both at low area densities and at high-$Z$.

\begin{figure}
\begin{centering}
\includegraphics[width=0.49\textwidth]{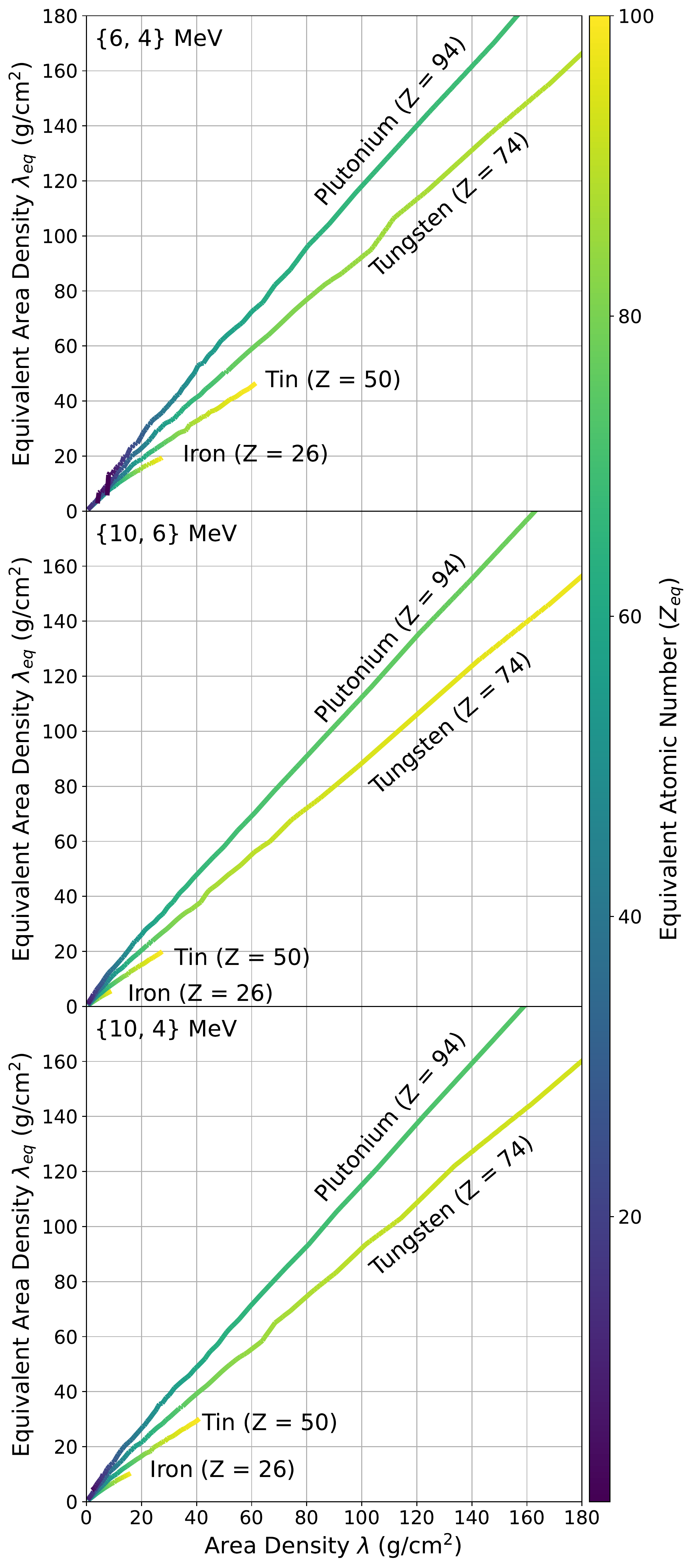}
\caption{Tracking the iron $(Z=26)$, tin $(Z=50)$, tungsten $(Z=74)$, and plutonium $(Z=94)$ $\alpha$-lines (Fig. \ref{alpha_curve}) and recording all $\alpha$-line intersections for dual energy $\{6, 4\}$ MeV (top), $\{10, 6\}$ MeV (center), and $\{10, 4\}$ MeV (bottom). At each $\alpha$-line intersection, the $\lambda$ of the tracked $\alpha$-line is recorded, along with the $\lambda_{eq}$ and $Z_{eq}$ of the intersected $\alpha$-line, which are then plotted to form an equivalent-line for each element. At these intersections, material $(\lambda, Z)$ and material $(\lambda_{eq}, Z_{eq})$ would produce the exact same transparencies, corresponding to two solutions to Eq. \ref{2x2eq}.}
\label{non_unique}
\end{centering}
\end{figure}

The results of this section are summarized in Fig. \ref{non_unique}, which shows equivalent-lines for various elements, identifying non-unique solutions to Eq. \ref{2x2eq} as area density is varied. At each point on the equivalent-line, the colorbar shows the element $Z_{eq}$ that, for area density $\lambda_{eq}$, would produce identical transparency measurements $\{T_H, T_L\}$ as the true material $Z$ at an area density of $\lambda$. Figure \ref{non_unique} reveals regions where atomic number discrimination capabilities are limited or impossible. In this analysis, the iron equivalent-line ends at $\lambda = 28 ~\g/\cm^2$ for the $\{6, 4\}$ MeV case, at $\lambda = 9 ~\g/\cm^2$ for the $\{10, 6\}$ MeV case, and at $\lambda = 16 ~\g/\cm^2$ for the $\{10, 4\}$ MeV case. Above this area density, iron can theoretically be uniquely characterized. However, below this area density, there exists a solution degeneracy in which a different pure material is fundamentally indistinguishable from iron. In real applications, the extent of this challenge may be significantly reduced through careful filtering of the bremsstrahlung beam~\cite{Ogorodnikov2002}.

At an intermediate area density of $\lambda = 100 ~\g/\cm^2$, the plutonium equivalent-line evaluates to $Z_{eq} = 66$ for the $\{6, 4\}$ MeV case, $Z_{eq} = 75$ for the $\{10, 6\}$ MeV case, and $Z_{eq} = 71$ for the $\{10, 4\}$ MeV case. These values for $Z_{eq}$ serve as the theoretical low-$Z$ bound of a high-$Z$ class. The degeneracy at high-$Z$ is still present even when using different bremsstrahlung beam filters. This property restricts the theoretical capabilities of high-$Z$ material identification to a high-$Z$ class, with atomic number discrimination within this class fundamentally ambiguous. It is important to clarify that this degeneracy at high-$Z$ is not solved by using dual monoenergetic beams, as the non-uniqueness property is due to a large photoelectric absorption cross section for high-$Z$ materials, not due to the polychromatic nature of the beams. Overall, the non-uniqueness challenges are most pronounced in the $\{6, 4\}$ MeV case, with the $\{10, 6\}$ MeV case outperforming the $\{10, 4\}$ MeV case.

\subsection{Sensitivity to Initial Measurements}
\label{ill conditioned}

As is frequently the case when solving inverse problems, the solution to Eq. \ref{2x2eq} is highly sensitive to the measured transparencies $T_H$ and $T_L$. This is significant because all measurements are inevitably corrupted by noise, meaning a poorly conditioned inversion will require high resolution measurements to obtain an accurate solution. The primary cause of the poor conditioning is due to Compton scattering being the dominant interaction mechanism within the bremsstrahlung beam energy range~\cite{Ogorodnikov2002}. Since the ratio $\frac{Z}{A}$ is approximately constant across different materials, $\mu_\CS(E, Z)$ is mostly insensitive to atomic number (Eq. \ref{muCS}).

One way to visualize the capabilities for atomic number discrimination is to consider the separation between different $\alpha$-lines on the $\alpha$-curves shown in Fig. \ref{alpha_curve}. $\alpha$-lines that are close together indicate a more challenging inversion since a purer signal is necessary to differentiate between different elements. The $\alpha$-curves presented in Fig. \ref{alpha_curve} show that the $\alpha$-line separation is greatest for lower-$Z$ elements and becomes smaller for higher-$Z$ elements. This reveals that atomic number discrimination becomes progressively more challenging as $Z$ increases due to the increased competition from the photoelectric effect.

The $\{10, 4\}$ MeV case shows the largest $\alpha$-line separation. This is due to the large contribution of pair production in the high energy $10$ MeV beam while maintaining minimal pair production in the low energy $4$ MeV beam. The $\{10, 6\}$ MeV case shows significantly worse $\alpha$-line separation than the $\{10, 4\}$ MeV case because of additional pair production in the low energy $6$ MeV beam. This reduces the contrast between the high and low energy beams, effectively diluting the atomic number dependence of the high energy $10$ MeV beam. The $\{6, 4\}$ MeV case shows the worst performance due to the reduced pair production in the high energy $6$ MeV beam. 

\subsection{Atomic Number Discrimination in the Context of Compound Materials}
\label{compound materials}

It has thus far been assumed that each material is composed entirely of a single element. In the case of heterogeneous materials, an effective atomic number $Z_\eff$ is assigned. Definitions of $Z_\eff$ vary and are frequently ambiguous. While it is possible to define the $Z_\eff$ of a compound through a weighted average of atomic numbers~\cite{Naydenov2013, Langeveld2017_effective_Z}, this definition is not so useful in the context of dual energy radiography. This is because the transparency measurements of a compound under this definition may differ from the transparency measurements of a homogeneous material with $Z = Z_\eff$, meaning perfect atomic number reconstruction would always differ from ground truth. Instead, the preferred definition for the $Z_\eff$ of a heterogeneous material is the atomic number that, if the entire material were composed of a single element with an effective area density $\lambda_\eff$, would produce the same high and low energy transparency measurements as the compound itself. For this reason, $Z_{\eff}$ is not an intrinsic property of a heterogeneous material, but rather dependent on the scanning system and thus different experiments will calculate different values for $Z_{\eff}$.

\begin{figure}
\begin{centering}
\includegraphics[width=0.49\textwidth]{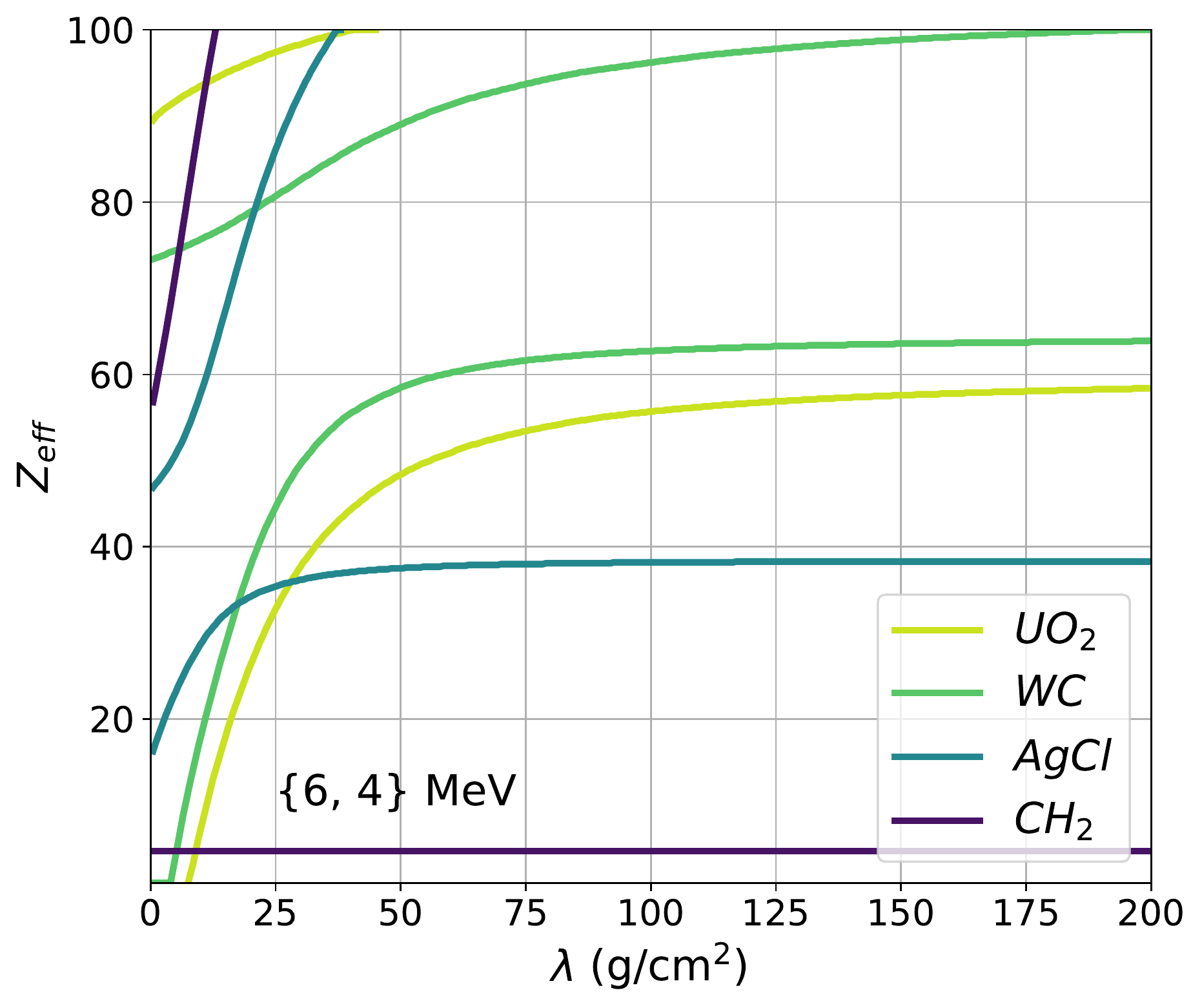}
\caption{The effective atomic number $Z_\eff$ of uranium oxide ($UO_2$), tungsten carbide ($WC$), silver chloride ($AgCl$), and polyethylene ($CH_2$) as a function of the area density $\lambda$ of the compound, generated using the $\{6, 4\}$ MeV analysis case. Hardening of the bremsstrahlung beams causes the $Z_\eff$ of compound materials to vary with material thickness. The presence of multiple lines for the same compound indicates non-unique solutions to Eq. \ref{2x2eq}.}
\label{compound}
\end{centering}
\end{figure}

Figure \ref{compound} shows the effective atomic number of various compound materials as a function of area density for the $\{6, 4\}$ MeV analysis case. Due to beam hardening, the $Z_\eff$ of a compound material is dependent on the material thickness. Furthermore, for the same reasons described in Section \ref{non-unique solutions}, the $Z_\eff$ of a compound material is not guaranteed to be unique. However, under most circumstances, the $Z_\eff$ of a compound material tends to be an intuitively intermediate atomic number. Overall, dual energy systems are limited in their ability to identify heterogeneous materials due to the nonspecificity of $Z_\eff$. A high-$Z$ nuclear material could be effectively hidden from dual energy radiography by surrounding it with low-$Z$ shielding or by embedding it within lower-$Z$ cargo contents. Nonetheless, a determination of $Z_\eff$ may still enhance cargo content analysis capabilities.

\section{Conclusion}

This analysis considers the capabilities and limitations for atomic number discrimination using dual energy $\{6, 4\}$ MeV, $\{10, 6\}$ MeV, and $\{10, 4\}$ MeV bremsstrahlung beams. In all three analysis cases, it is fundamentally impossible to uniquely characterize different materials at low area densities and at high-$Z$. This property stems from competition between photoelectric absorption and pair production, whereby different pure materials can produce identical transparency measurements. This result creates a fundamental ambiguity when discriminating between materials of different atomic number, even when using a system with perfect statistical resolution. While unique high-$Z$ material identification in cargo is impossible using currently deployed commercial dual energy bremsstrahlung scanners, a high-$Z$ material class is possible in principle. Future research should use commercial radiographic systems to experimentally validate these conclusions. These efforts should also study whether computed tomography techniques may mitigate these limitations

Given the multiple fundamental problems raised by this study, future research in the field of cargo security should also consider the capabilities of active interrogation techniques, which would produce signal terms that are unique to fissile special nuclear materials. Past research in the area has shown that techniques such as prompt neutrons from photofission, nuclear resonance fluorescence, and other methodologies hold significant promise in resolving the above-described non-uniqueness inherent to simple radiographic techniques. For some examples of these see Ref.~\cite{pnpf-short,NRF_bertozzi,stevenson2011linac,runkle_rattling}.

\section{Acknowledgements}

This work was supported by the Department of Energy Computational Science Graduate Fellowship (DOE CSGF) under grant DE-SC0020347. The authors acknowledge advice from Brian Henderson for his expertise on the technical nuances of dual energy radiography systems. The authors declare no conflict of interest.
\newpage

\onecolumn

\bibliography{References.bib}

\twocolumn

\newpage

\appendix

\section{}
\label{appendix}

\subsection{Simulation of the Beam Spectra and Detector Response}
\label{Simulation}

\begin{figure}[t]
\begin{centering}
\includegraphics[width=0.49\textwidth]{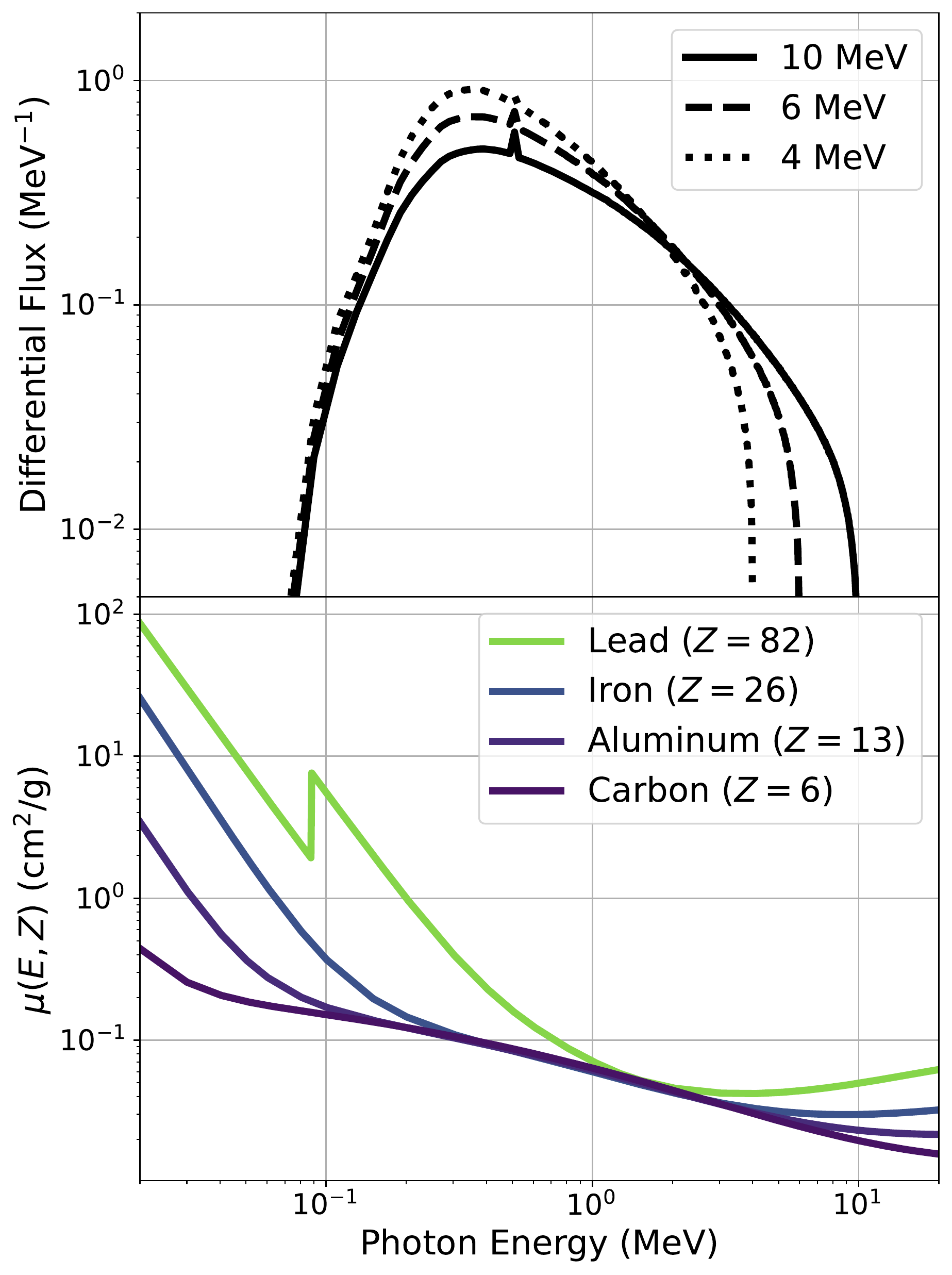}
\caption{Simulated $4$ MeV, $6$ MeV, and $10$ MeV bremsstrahlung beam spectra $\phi(E)$ (top) along with the mass attenuation coefficients for carbon, aluminum, iron, and lead (bottom). Compton scattering is the dominant interaction mechanism within the bremsstrahlung energy range.}
\label{spectra}
\end{centering}
\end{figure}

Simulations of the incident beam spectra $\phi_{\{H, L\}}(E)$ and detector response function $D(E)$ were run in Geant4~\cite{Geant4, grasshopper}. Simulations were performed using the QGSP BIC physics list with a production cut of 80 keV. The geometries were designed to be simple, generalizable, and representative of typical cargo scanners. In order to model the bremsstrahlung beam spectra, electrons were directed at a 0.1cm tungsten radiator backed by 1cm of copper, with an additional 1cm of steel to filter the low energy X-rays. $4$ MeV electrons were used when simulating the $4$ MeV endpoint energy bremsstrahlung spectrum, and similarly for the $6$ and $10$ MeV endpoint energy bremsstrahlung spectra. The bremsstrahlung photons were then recorded by a tally surface $(r = 0.2\cm)$ placed 10cm behind the tungsten target, subtending a half-angle of 20 milliradians $(\approx 1.15^{\circ})$. The resulting beam spectra are shown in Fig. \ref{spectra}. The detector response function was calculated by building a $1.5 \times 1.0 \times 3.0 \cm$ cadmium tungstate (CdWO$_4$) crystal. Photons with energy uniformly distributed between $0$ and $10$ MeV were then directed along the long axis of the crystal. For each photon incident on the detector, the total energy deposited in the detector was recorded. This result was then binned by the incident photon energy to produce the detector response function.

\subsection{Transparency Simulations}
\label{Transparency Simulations}

\begin{figure}
\begin{centering}
\includegraphics[width=0.49\textwidth]{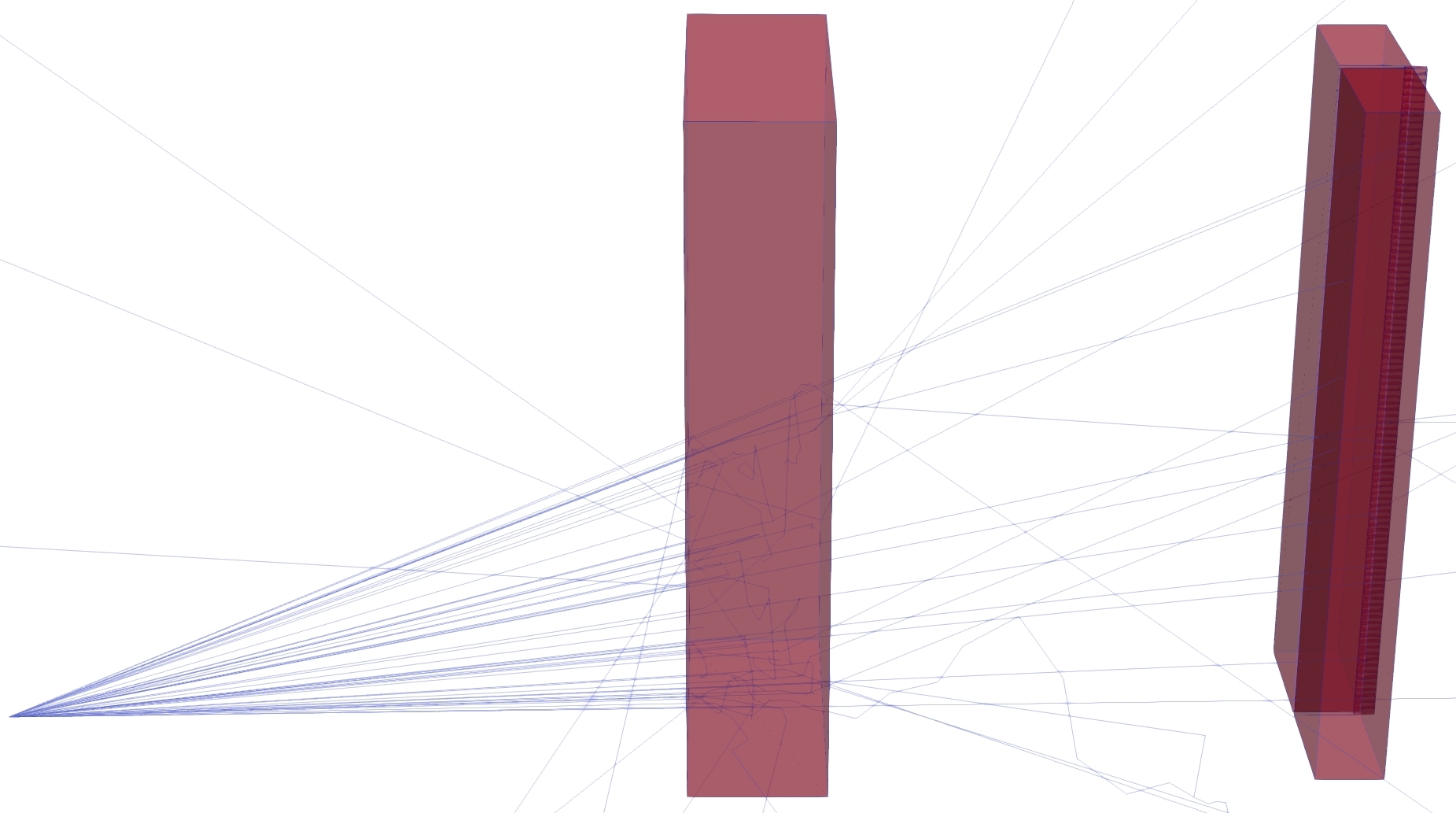}
\caption{A bremsstrahlung fan beam (left) is directed towards a target (center). A collimated stack of CdWO$_4$ detectors (right) measures the transparency of the photon beam.}
\label{simGeom}
\end{centering}
\end{figure}

To simulate transparency measurements of a target material, photons with energy sampled from $\phi_H$ or $\phi_L$ were directed in a fan beam through a large target box. The photons were then detected by a stack of CdWO$_4$ detectors operating in energy-integrating mode. A $10$cm lead collimator was used to filter scattered radiation. This simulation was repeated for different target area densities and atomic numbers to obtain a simulation dataset. The simulation geometry is shown in figure \ref{simGeom}.

\subsection{Calculation of the Calibration Parameters}
\label{calculating_abc}

In radiographic systems, the presence of scattering causes the detected transparency to diverge from an idealized free streaming photon model. Our past work showed that these secondary effects can be accounted for by substituting a semiempirical mass attenuation coefficient (Eq. \ref{semiempirical_mass_atten}) into the analytic transparency model (Eq. \ref{transparency})~\cite{Lalor2023}. To calibrate the model, high resolution transparency simulations of carbon $(Z=6)$, iron $(Z=26)$, and lead $(Z=82)$ were performed at an area density of $\lambda = 150 \g/\cm^2$. Then, $a$, $b$, and $c$ are determined by minimizing the squared logarithmic error between the transparency model and transparency simulations:

\begin{equation}
a, b, c = \argmin_{a', b', c'} \sum_i \left(\log T(\lambda_i, Z_i) - \log T_i\right)^2
\label{calc_abc}
\end{equation}

where $T_i$ is the simulated transparency of calibration material $\{\lambda_i, Z_i\}$. This step was performed separately for the $10$, $6$, and $4$ MeV bremsstrahlung beams. The values for the calibration parameters used in this study are shown in table \ref{abc_fit}.

\begin{table}
\begin{centering}
\begin{tabular}{c c c c}
\toprule
Endpoint energy & a & b & c \\
\midrule
10 MeV & 1.1113 & 0.9924 & 0.9861 \\
6 MeV & 0.9677 & 0.9889 & 0.9921 \\
4 MeV & 0.9143 & 0.9883 & 0.9945 \\
\bottomrule
\end{tabular}
\caption{Calibration parameters $a$, $b$, and $c$ for each of the analysis cases considered in this study.}
\label{abc_fit} 
\end{centering}
\end{table}

\subsection{Thickness Equivalent Conversion}

\begin{figure}
\begin{centering}
\includegraphics[width=0.49\textwidth]{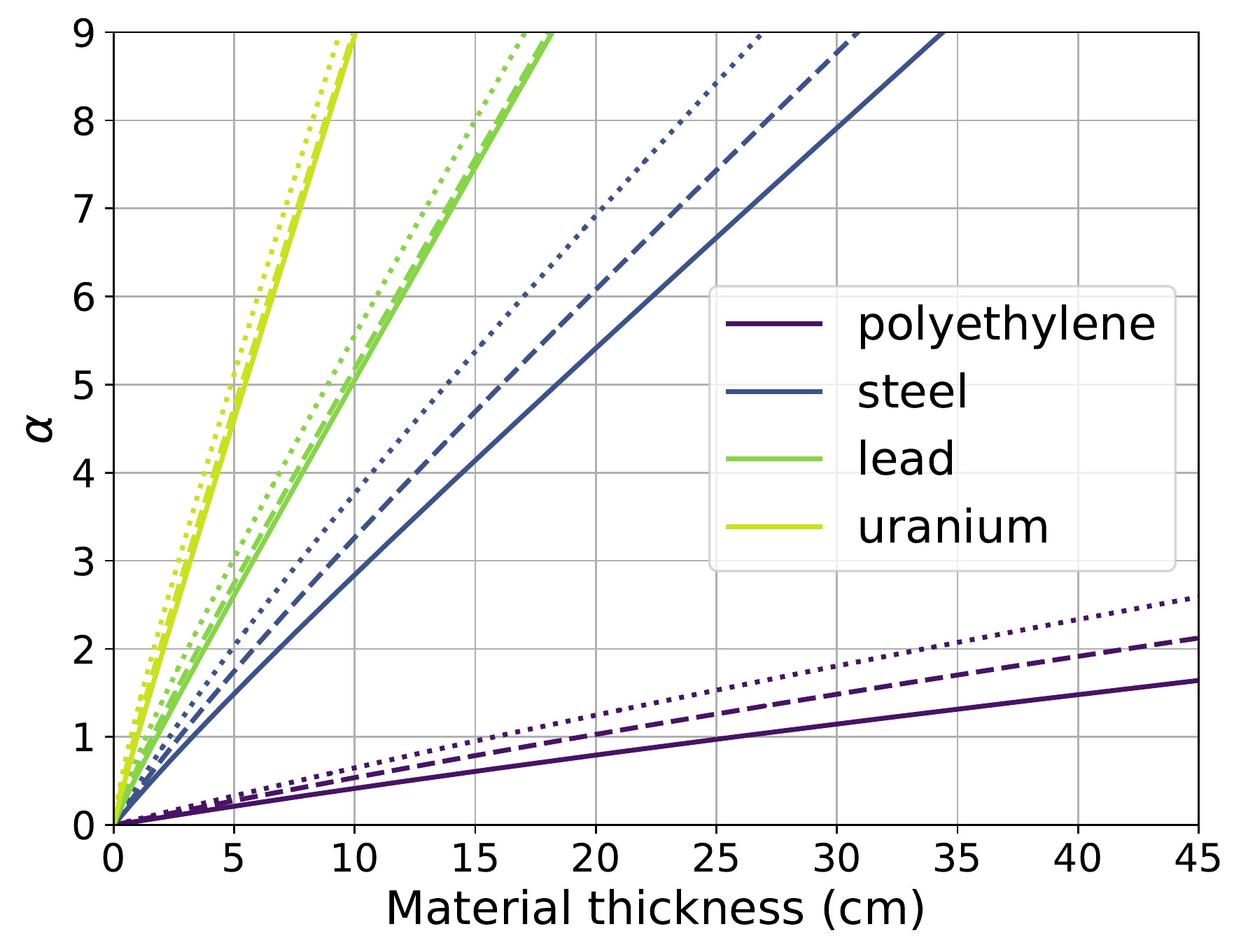}
\caption{The log-transparencies (Eq. \ref{alpha}) produced by different thicknesses of polyethylene $(\rho = 1 ~\g/\cm^3)$, steel $(\rho = 8 ~\g/\cm^3)$, lead $(\rho = 11 ~\g/\cm^3)$, and uranium $(\rho = 19 ~\g/\cm^3)$ when using a $10$ MeV bremsstrahlung beam (solid line), $6$ MeV bremsstrahlung beam (dashed line), and $4$ MeV bremsstrahlung beam (dotted line). $\alpha$ is roughly proportional to material thickness, although due to beam hardening, the proportionality is not exact.}
\label{x_eq}
\end{centering}
\end{figure}

Using Eq. \ref{alpha}, the log-transparency can be calculated for a given material $Z$ and area density $\lambda$ in the presence of an incident beam spectrum $\phi(E)$. In order to physically interpret the particular values of $\alpha$, Fig. \ref{x_eq} shows $\alpha$ as a function of material thickness for different materials for a range of thicknesses. For reference, $\alpha = 9$ corresponds to approximately $31 \cm$ of steel using a $6$ MeV bremsstrahlung beam, which is the maximum penetration of the Rapiscan Eagle R60${}^\text{\textregistered}$ scanning system~\cite{Rapiscan_maxPen}

\end{sloppypar}
\end{document}